\documentclass[twocolumn,preprintnumbers,amsmath,amssymb]{revtex4}
\usepackage{graphicx}
\usepackage{amsmath}
\usepackage{amssymb}
\usepackage{dcolumn}
\usepackage{color}
\usepackage{epstopdf}

\newcommand{\zetacond}{\zeta^{(c)}}
\newcommand{\zetathermal}{\zeta^{(th)}}

\begin{document}

\title{Long timescale dynamics of spin textures in a degenerate F=1 $^{87}$ Rb spinor Bose gas }
\author{J. Guzman$^1$}
\author{G.-B. Jo$^1$}
\author{A. N. Wenz$^{1,3}$}
\author{K. W. Murch$^{1}$}
\author{C. K. Thomas$^{1}$}
\author{D. M. Stamper-Kurn$^{1,2}$}
\affiliation{
    $^1$Department of Physics, University of California, Berkeley CA 94720 \\
    $^2$Materials Sciences Division, Lawrence Berkeley National Laboratory, Berkeley, CA 94720\\
    $^3$Physikalisches Institut, Ruprecht-Karls-Universit\"{a}t Heidelberg, 69120 Heidelberg, Germany}

\date{\today }

\begin{abstract}

We investigate the long-term dynamics of spin textures prepared by cooling unmagnetized spinor gases of $F=1$ $^{87}$Rb to quantum degeneracy, observing domain coarsening and a strong dependence of the equilibration dynamics on the quadratic Zeeman shift $q$.  For small values of $|q|$, the textures arrive at a configuration independent of the initial spin-state composition, characterized by large length-scale spin domains, and the establishment of easy-axis (negative $q$) or easy-plane (positive $q$) magnetic anisotropy.  For larger $|q|$,  equilibration is delayed as the spin-state composition of the degenerate spinor gas remains close to its initial value. These observations support the mean-field equilibrium phase diagram predicted for a ferromagnetic spinor Bose-Einstein condensate, but also illustrate that equilibration is achieved under a narrow range of experimental settings, making the $F=1$ $^{87}$Rb gas more suitable for studies of nonequilibrium quantum dynamics.


\end{abstract}


\maketitle

Extensive efforts are underway to utilize ultracold atoms as simulators of condensed-matter systems, making use of their broad tunability and amenity to novel experimental probes \cite{bloc08rmp}.  The success of these endeavors hinges on the cold-atom system's arriving at equilibrium, matching the conditions of solid state systems that reside at or near the thermal equilibrium state.  However, given their limited lifetime, weak interactions, long dynamical timescales, and several nearly conserved quantities, whether equilibrium is reached in cold-atomic materials has become a pressing scientific question \cite{kino06,hung10slowmass,natu11local}.

One promising system for studying materials physics is the spinor Bose gas, comprised of atoms that may explore all Zeeman states within a spin $F$ manifold.  At low temperature, quantum degenerate spinor gases exhibit spontaneous symmetry breaking \cite{sadl06symm,sche10}, magnetic order, and intricate spin textures \cite{veng08helix,veng10periodic} governed by the interplay of magnetism and superfluidity. Several works have examined the ground state of a spatially uniform $F=1$ spinor gas for which the spin-dependent contact interaction, characterized by a coupling strength $c_2 = 4 \pi \hbar^2 \Delta a / m < 0$, favors magnetized spin states \cite{ho98,ohmi98,law98spin2,sten98spin}; here $\Delta a$ is a difference in $s$-wave scattering lengths, and $m$ is the atomic mass.  Under the application of a quadratic Zeeman energy $q F_x^2$, with $F_x$ being the dimensionless spin projection along an experimentally selected axis $\hat{x}$, and neglecting linear Zeeman shifts, both mean-field and exact many-body treatments predict a symmetry breaking phase transition between paramagnetic and ferromagnetic phases at a positive quadratic Zeeman shift $q = q_0 = 2 |c_2| n$ where $n$ is the gas density.  Within the ferromagnetic phase, the magnetization is favored to lie transverse to $\hat{x}$ for $q>0$ (easy-plane) and along $\hat{x}$ for $q<0$ (easy-axis).

This predicted phase diagram (for $q>0$) was supported by studies of  $F=1$ spinor condensates of $^{87}$Rb in the single spatial-mode regime, for which the condensate dimensions in all directions were comparable to the spin healing length, defined by $\xi_s = (8 \pi \Delta a n)^{-1/2}$ \cite{schm04,chan04,chan05nphys}.  However, recent experiments on spatially extended spinor gases revealed variegated magnetization textures produced upon cooling unmagnetized spinor gases into the quantum degenerate regime \cite{veng10periodic}.  Theoretical works examined whether such textures could be favored at equilibrium by the substantial magnetic dipolar interactions in the $^{87}$Rb gas; however, while some inhomogeneous textures do appear energetically favored, or at least metastable, they are predicted to evince spin-density modulations only on much longer lengths scales than observed experimentally \cite{taka07classical,cher09roton,kjal09,kawa10spont,zhan10jltp,huht10}.

To resolve this discrepancy, we examine quantum-degenerate spinor Bose gases over evolution times of several seconds, much longer than previously \cite{veng10periodic}.  We find that cooling non-degenerate gases with different initial spin compositions produces spin textures that differ strongly at short equilibration times, consistent with Ref.\ \cite{veng10periodic}.  However, at least at small values of $|q|\lesssim h \times 10 \, \mbox{Hz}$ and at the longest equilibration times, these textures do tend toward a common state characterized by large commonly magnetized regions and easy-plane/easy-axis magnetic anisotropy consistent with the predicted equilibrium phases of ferromagnetic spinor condensates with contact interactions.  For larger values of $|q|$, the redistribution of Zeeman populations needed to yield the expected equilibrium states is dramatically slowed, preventing the appearance of a common state from differing initial populations.

For our experiments, we produce optically trapped, spin-polarized atomic gases in a manner similar to Ref.\ \cite{lin09rapid}, with initial temperatures of 30 $\mu$K and initial trap frequencies of $(\omega_x,\omega_y,\omega_z)=2\pi(220,4150,63)$ Hz in a focused, linearly polarized light beam with a wavelength of 1064 nm.  We next prepare unmagnetized spin mixtures with fractional populations in the three Zeeman sublevels, 
 $\zeta^{(th)}_{m_F}$ with $m_F= \{+1,0,-1\}$, by the application of resonant $\pi/2$ rf pulses in the presence of a magnetic field gradient \cite{veng10periodic}. The spin mixtures are then cooled by lowering the optical trap depth over 2.4 s to a final trap depth of $1.5\,\mu$K and final trap frequencies of $(\omega_x,\omega_y,\omega_z)=2\pi(25,480,7.3)$ Hz. We then allow for an equilibration period of up to $t=4$ s, where $t=0$ denotes the end of the ramp-down of the optical trap depth.

Throughout the evaporation and equilibration periods, a static magnetic field of $B=267$ mG was applied along the $\hat{x}$ axis, with a spatial inhomogeneity of less than 1 $\mu$G across the extent of the gas. This field causes the transverse magnetization of the sample to precess with respect to the fixed trap axes at the Larmor frequency of $\Omega_L = 2 \pi \, \times \, 187$ kHz.  The field also produces a quadratic Zeeman energy shift $q_B=h (70\,\text{Hz/G}^2) B^2\approx h\times 5$ Hz.  The sample was also exposed to a linearly polarized microwave-frequency magnetic field detuned by $\delta = \pm 2 \pi \times 40$ kHz from the $\vert F=1, m_F = 0\rangle$ to $\vert F=2, m_F = 0\rangle$ hyperfine transition.  The resulting ac Zeeman shift of $q_\mu = - \hbar \Omega^2 / 4 \delta$  was used to tune the total quadratic shift $q = q_B + q_\mu$ by varying the Rabi frequency $\Omega$ and by choosing the sign of $\delta$ \cite{lesl09amp,gerb06rescontrol}.

The spinor gas was then measured either by time-of-flight analysis, through which we quantify the fractional Zeeman populations, $\zetacond_m$, of the "quasi-condensed" portion of the gas (defined conventionally by the bimodality of the spatial distribution of the expanded gas), or by \emph{in-situ} measurements of the vector magnetization.  We first discuss  the time-of-flight measurements, which were performed on gases that were prepared initially with fractional Zeeman populations $\zeta^{(th)}_{m_F}$ of either 0, $1/4$ or $1/3$.  

\begin{figure}[!h]
\centering
\includegraphics[width=0.45\textwidth]{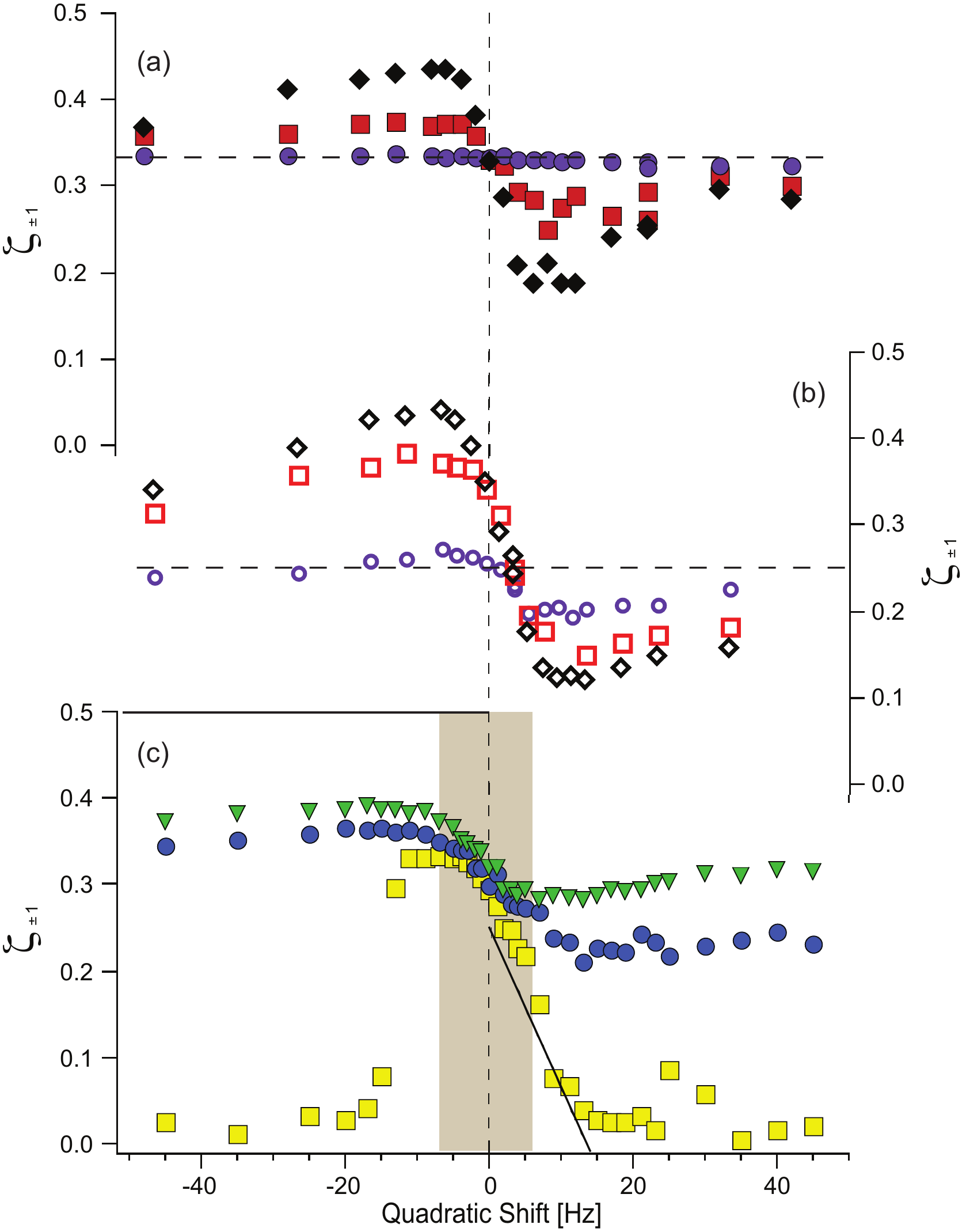}
\caption{Condensate Zeeman populations, $\zetacond_{\pm 1}$, during equilibration of unmagnetized spinor gases at variable $q$.  Measurements at various evolution times $t=$ 200 ms (purple circles), 1 s (red squares), and 3 s (black diamonds) are shown with initial $\zetathermal_{\pm 1} = 1/3$ (a) and $1/4$ (b).  In (c), Zeeman populations at $t = 2$ s are compared for initial $\zetathermal_{\pm 1} = 1/3$ (green triangles), $1/4$ (blue circles), and $0$ (yellow squares).  For a narrow range $|q|/h\lesssim 5$ Hz (shaded), Zeeman populations evolve toward a common steady state within about 2 s, toward values qualitatively in accord with mean-field predictions (solid line in c) based on the spin-dependent contact interaction.  Outside that range, initial population differences persist throughout the experimentally accessed equilibration times.  Data are averages of 3 experimental runs.} \label{fig:tof}
\end{figure}

Evaporatively cooling such gases produces spinor condensates containing around $3 \times 10^6$ atoms at $t=0$, and around $1.2 \times 10^6$ atoms at $t=4$ s.  Assuming a conventional Thomas-Fermi distribution of condensed atoms within the optical trap, we determine the peak condensate densities to be $n_0=2.6 \times 10^{14}\,  \mbox{cm}^{-3}$ at  $t=0$ and 1.8 $\times 10^{14}\, \mbox{cm}^{-3}$ at $t=4$ s, respectively. We note that bimodal time-of-flight distributions are observed already beginning at $t \simeq -50$ ms.  At early times the spinor condensates are found to have fractional Zeeman populations identical to those of non-degenerate gas from which the condensates form.  Following that, spin-mixing collisions within the condensate cause the Zeeman populations to vary.  For instance, starting with equal populations in the three Zeeman populations Fig.\  \ref{fig:tof}(a), the $\zetacond_{\pm 1}$ population tends to decrease when the energy of the $\vert m_F = \pm 1 \rangle$ states is raised ($q>0$), and to increase when it is lowered ($q<0$). Comparing measurements on samples with different initial spin compositions (Fig.\  \ref{fig:tof}(c)), we find that a common steady-state distribution among Zeeman levels is achieved within several seconds of equilibration, for $|q|\lesssim10$ Hz.

\begin{figure*}[!t]
\centering
\includegraphics[width=1.0\textwidth]{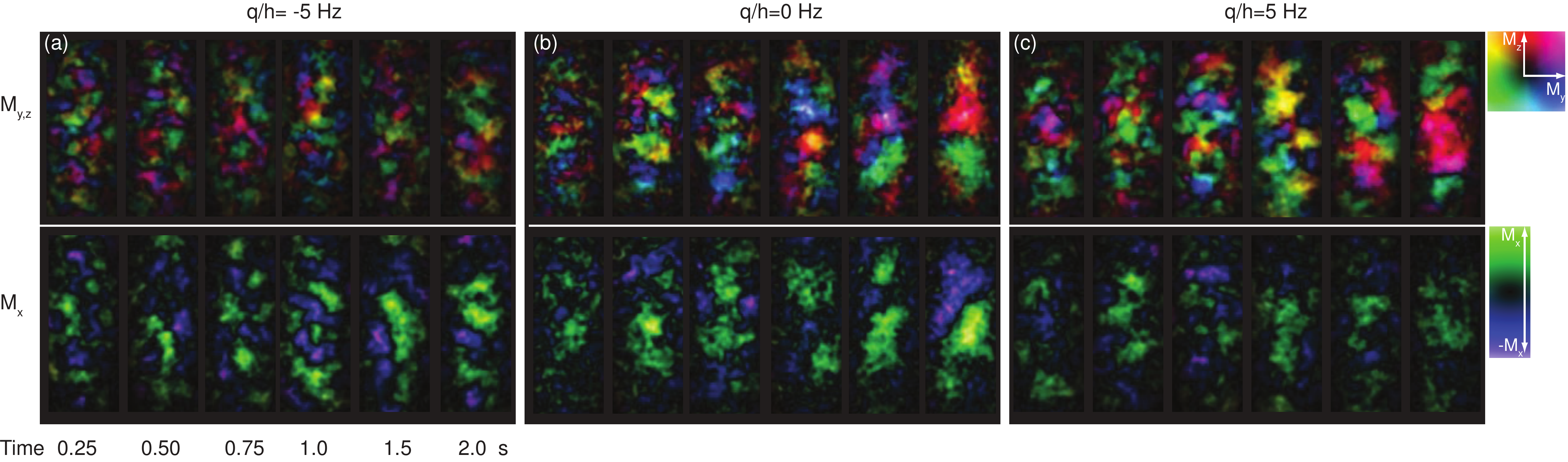}
\caption{Transverse (top) and longitudinal (bottom) magnetization after a variable evolution time at a quadratic shift of $q/h=$ -5 (a), 0 (b), and 5 (c) Hz for an initial spin mixture of $\zetathermal_{\pm 1}=1/3$. The transverse magnetization is represented using the color scheme shown, where the hue indicates the magnetization orientation and the brightness its magnitude.  The maximum brightness is set by a a fully polarized condensate at 2 s evolution time. 
The longitudinal magnetization is represented by the color bar. Textures form from initially small domains and then coarsen to where only a few domains span the entire condensate. The late-time images show a spin-space anisotropy;  for positive $q$ the transverse magnetization appears brighter while for negative $q$ the opposite trend is observed.} \label{fig:PCvstime}
\end{figure*}

We compare the measured Zeeman populations to the mean-field predictions for the uniform ground state of a ferromagnetic $F=1$ spinor condensate \cite{sten98spin}.  For $q<0$, the spin-dependent contact interaction is minimized by a fully magnetized condensate with magnetization aligned with the field axis $\hat{x}$.  For our experiments, the total magnetization $\langle F_x \rangle = 0$ is conserved in the absence of dipolar relaxation.  To satisfy this constraint, one might expect the condensate to be divided into domains of magnetization oriented along $\pm \hat{x}$, resulting in $\zetacond_{\pm 1} = 0.5$.  For $0<q<2 |c_2| \langle n\rangle_y \simeq h \times 8$ Hz,  where $\langle n \rangle_y=1.85\times 10^{14}\, \mbox{cm}^{-3} $ is the density averaged in the $\hat y$ direction after 2 s of equilibration, one expects that the condensate comprise transversely polarized spin domains in which $\zetacond_{\pm 1}$ diminishes monotonically from 0.25 to zero \cite{note:zeta1}.  At $q = 0$, fully magnetized domains of all spin orientations minimize the mean-field energy; one might therefore expect $\zetacond_{\pm 1} = 1/3$.  These predictions ignore the number and composition of domain boundaries, and agree qualitatively with our measurements at small $|q|$.

For larger values of $|q|$, the Zeeman populations do not reach a common steady state within the 4 s lifetimes of our trapped samples.  For $q > 2 |c_2| n_0$, the single-atom quadratic Zeeman shift is predicted to be sufficient to overcome the ferromagnetic spin-dependent contact interaction and lead to a condensate equilibrating purely in the $\vert m_F = 0\rangle$ state \cite{sten98spin,schm04,chan04}.  However, the condensates prepared with $\zetacond_{\pm 1} \neq 0$ retain non-equilibrium Zeeman populations nearly equal to their initial values, with the rate of Zeeman-population transfer diminishing with increasing $q$.  For $q<0$, the single-atom quadratic Zeeman shift favors population in the $\vert m_F = \pm 1\rangle$ states.  However, for $q/h \lesssim -20$ Hz, while condensates within non-zero initial values of $\zetathermal_{\pm 1}$ do evolve toward the presumed equilibrium distribution, a condensate prepared in the non-equilibrium $\vert m_F = 0\rangle$ Zeeman state shows almost no variation in its spin distribution.  The extinction of spin-mixing dynamics at negative $q$ for condensates in the $\vert m_F = 0 \rangle$ states was observed previously \cite{lesl09amp}.

The strong variation of spin-mixing dynamics with variations in $q$ that are far smaller than the thermal energy ($k_B (50 \, \mbox{nK}) = h \times 1$ kHz) suggests that spin-mixing occurs primarily within the condensed fraction of the gas, with little influence of the non-condensed portion.  The variation of condensate spin-mixing dynamics with $q$ has been studied in the single-mode regime \cite{chan05nphys,kron06tune}, showing that the amplitude of Zeeman-population oscillations diminishes at large quadratic shifts.  This reduction in amplitude occurs once the energy difference introduced between the two-atom states coupled by spin mixing becomes larger than the spin-mixing interaction strength ($|q|> 2 |c_2| n$).  However, it is unclear how this single-mode explanation applies to the inhomogeneous spin textures studied here.


While the Zeeman-population measurements are indicative of equilibration dynamics, they do not reveal directly the spontaneous magnetization of the quantum degenerate gas nor do they characterize the spatial distribution of the condensate spin.  To explore these properties, we employ magnetization-sensitive dispersive imaging.  In contrast to previous work \cite{higb05larmor,veng07mag}, in which the magnetization was reconstructed from a series of images taken during Larmor precession, here we obtain a direct snapshot of the vector magnetization with just three images by applying spin-echo rf pulses between probes.  Specifically, after a variable equilibration time, we apply the first imaging pulse, sending linear polarized probe light along the $\hat{y}$ axis and measuring its optical rotation to resolve one transverse component of the column-integrated magnetization $\mathbf{\tilde{M}}(\mathbf{\rho})$ with the position vector $\mathbf{\rho}$ in the imaged $\hat{x}$-$\hat{z}$ plane.  Following a delay $\tau_0$ of about 50 $\mu$s, we apply a $\pi$ pulse with a resonant rf magnetic field.  A second image, taken after an additional delay $\tau=\tau_{0}+ \pi/2 \Omega_L$, records the second component of the transverse magnetization.  To measure the longitudinal magnetization component, we apply a properly timed $\pi$-$\pi/2$-$\pi$ pulse sequence before taking a third image.  This spin-echo imaging sequence functions reliably even in the presence of $>2\pi \times 10$ kHz variations in $\Omega_L$ between experimental runs.  For future work, the relative parsimony of this method should allow either higher-resolution or repeated imaging of atomic spin distributions.

The temporal evolution of spin textures produced upon cooling thermal mixtures with $\zetathermal_{\pm 1} = 1/3$ is shown in Fig.\ \ref{fig:PCvstime} for three different settings of $q$.  Within several hundred milliseconds, a highly corrugated spin texture develops within the condensate consisting of small domains with magnetization orientations distributed isotropically.  The domains have a typical diameter of about 10 microns, in agreement with previous observations \cite{veng10periodic}, although the textures in the present experiment do not show a preferred direction for spatial modulation of the  magnetization in the $\hat{x}$-$\hat{z}$ plane.


The next few seconds of equilibration lead to distinct changes in both the composition and the spatial arrangement of the spin textures. The magnetization develops a $q$-dependent spin-space anisotropy visible in the contrasting magnetization amplitudes for different quadratic shifts (Fig. \ref{fig:PCvstime}). Also evident is the development of larger domains of common magnetization.  At early times the magnetization map consists of many randomly oriented spin domains with N\'eel-type domains walls, separating regions of common magnetization.  Subsequently, within 2-3 s, the texture contains just a few domains separated by N\'eel-type domains walls. 

To characterize these changes we consider the dimensionless correlation functions of the $i$ component of the magnetization vector, given as
\begin{equation}
G_i(\delta \mathbf{\rho}) = \frac{\sum_{\mathbf{\rho}} \tilde{M}_i(\mathbf{\rho} + \delta \mathbf{\rho}) \tilde{M}_i(\mathbf{\rho})}{\mu^2 \sum_{\mathbf{\rho}} \tilde{n}(\mathbf{\rho} + \delta \mathbf{\rho})\cdot \tilde{n} (\mathbf{\rho})},
\end{equation}
with $\mu$ being the atomic magnetic moment.  This correlation function is evaluated over the central 35 $\times$ 85 $\mu \text{m}^2$ area of the condensate.

\begin{figure}[!h]
\centering
\includegraphics[width=0.45\textwidth]{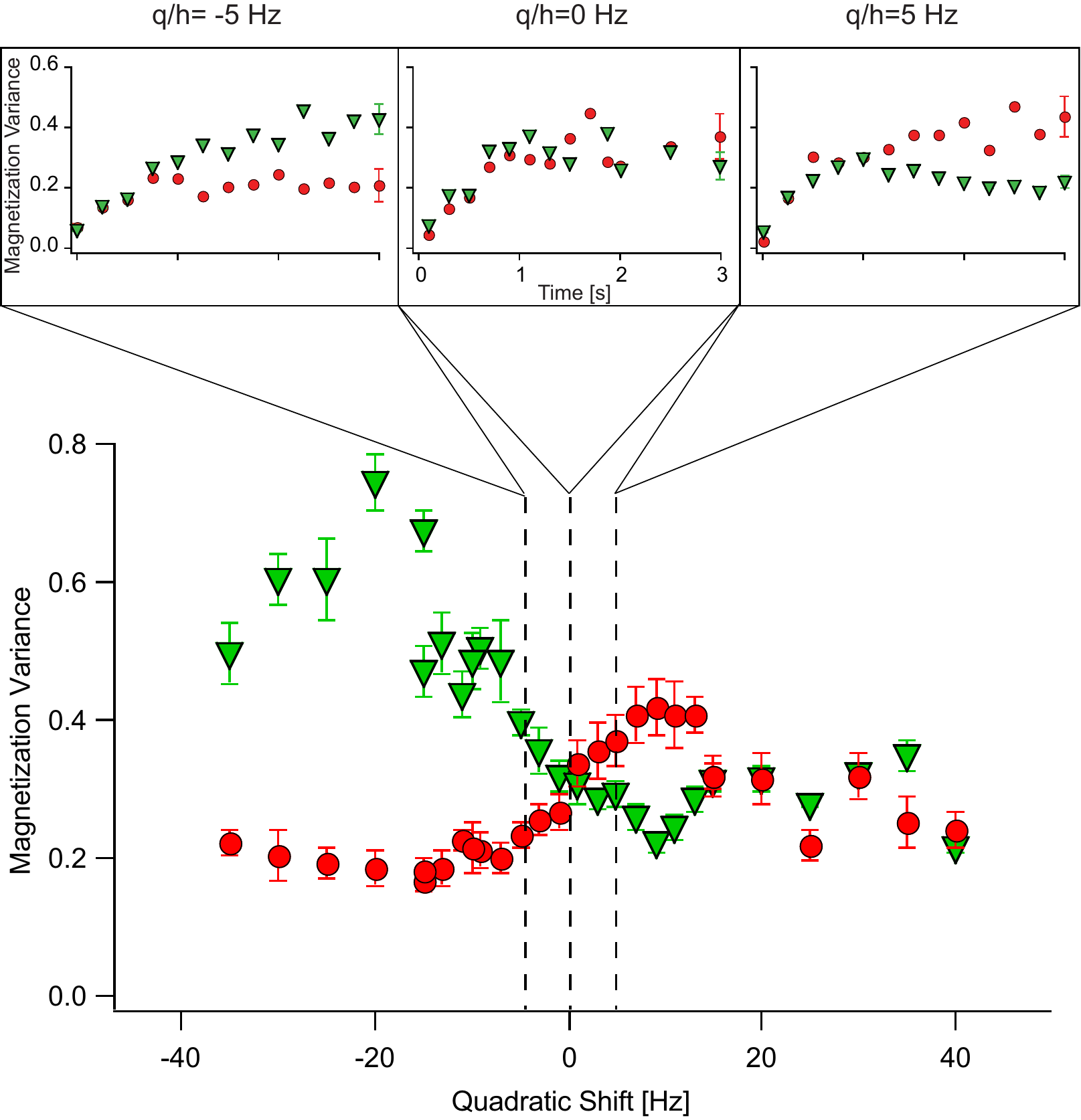}
\caption{The magnetization variance versus $q$ for $\zetathermal_{\pm 1}=1/3$ at t=2 s for transverse magnetization $G_T(0)$(red circles) and longitudinal magnetization $G_L(0)$ (green triangles). For $q=0$ the longitudinal and transverse magnetization variances are equal, showing no preference in spin orientation (isotropic), while for $q>0$ ($q<0$), the data show a preference for transverse (longitudinal) magnetization, seen in the larger transverse (longitudinal) magnetization variance (easy plane/easy axis). (Inset) Temporal evolution of the transverse ($G_T(0)$, red circles) and longitudinal ($G_L(0)$, green diamonds) magnetization variance, evaluated within the central regions of  spin textures produced from  $\zetathermal_{\pm 1}=1/3$ spin mixtures at $q/h= -5$ Hz (left), 0 Hz (middle) and 5 Hz (right). At early equilibration times, $t\lesssim 700$ ms, the magnetization has equal variance in the longitudinal and transverse orientations.  Within 1-2 s, a clear preference for longitudinal orientation at  $q>0$ (middle), while for $q=0$ the spin distribution remains isotropic.  Data are the average of 6 experimental repetitions, and error bars are the standard deviation of the data.} \label{fig:Variance}
\end{figure}

Setting $\delta \mathbf{\rho} = 0$, we obtain the variance of the longitudinal magnetization $G_L(0)$=$G_x(0)$, and also the average of the transverse magnetization $G_T(0)=(G_y(0)+G_z(0))/2$ \cite{note:variance}. For a fully magnetized gas $G_L(0)+2\times G_T(0)=1$.  As shown in Fig.\ \ref{fig:Variance} (inset) for an initially unpolarized gas with an initial spin composition of $\zeta^{(th)}_{\pm 1}=1/3$, magnetization at early equilibration times ($t \lesssim 700$ ms) develops with equal variance in the longitudinal and transverse orientations.  Thereafter, within 2 s, the sum of the variances are consistent with unity and the magnetization develops a spin-space anisotropy: easy-plane or easy-axis anisotropy seen from the enhanced transverse or longitudinal magnetization, for $q>0$ or $q<0$, respectively (Fig. \ref{fig:Variance}).  At $q=0$, the magnetization orientation remains isotropic.  Both the $q$-dependence and also the 1-2 s timescale for development of the spin-space anisotropy are consistent with the Zeeman-population measurements discussed above.

\begin{figure}[!h]
\centering
\includegraphics[width=0.45\textwidth]{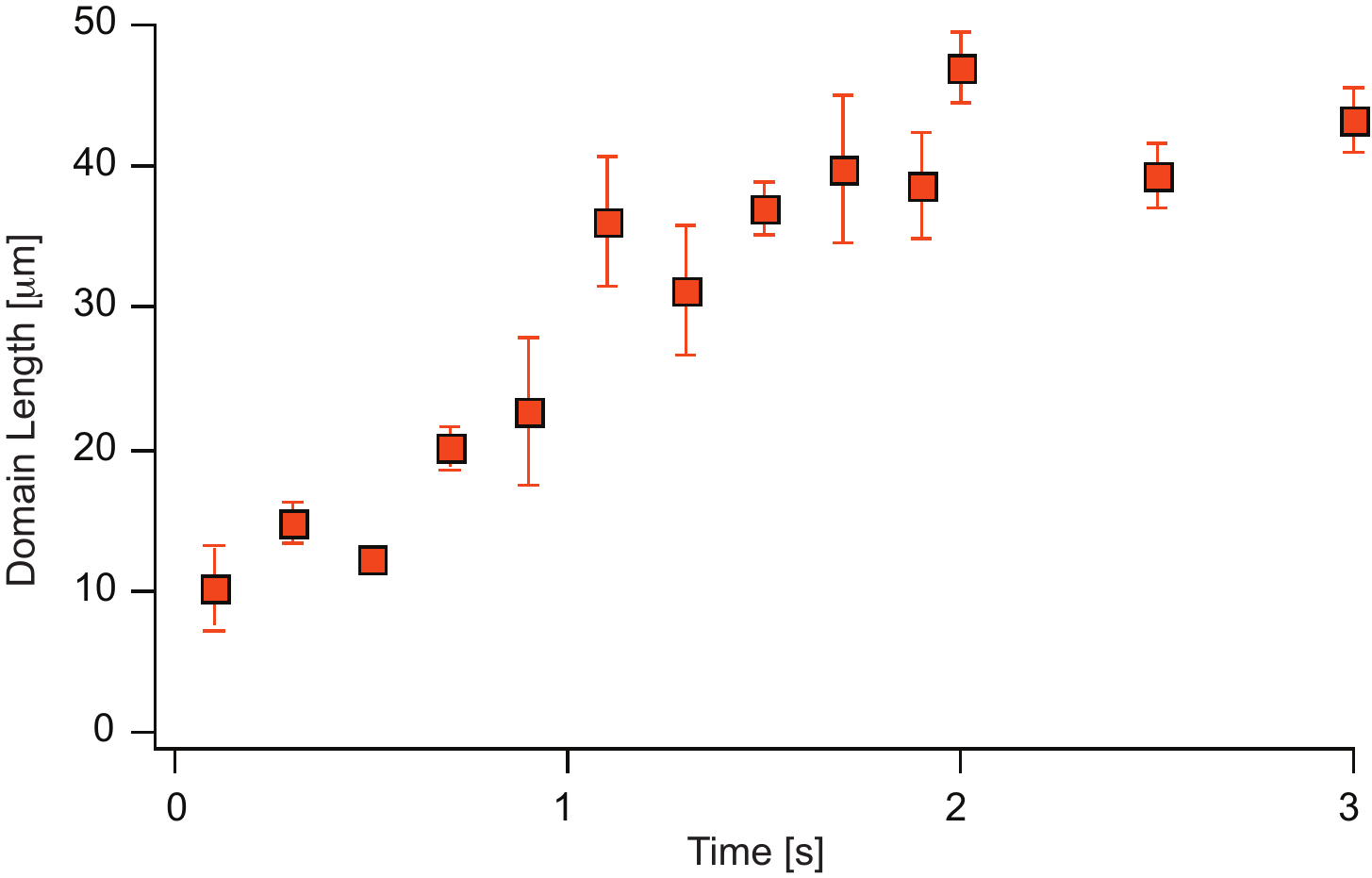}
\caption{Temporal evolution of domain length after variable evolution time at a quadratic shift of 0 Hz for an $\zeta^{th}_{\pm 1}=1/3$ mixture.  The domain length, as defined in the text, grows smoothly as a function of the evolution time, with domains of about 10 $\mu$m to final values of 40 $\mu$m within 2 s. } \label{fig:domainarea}
\end{figure}

A second trend in the evolution of these spin textures is the development of ever-larger domains of common magnetization.  The coarsening dynamics are characterized by measuring the area $A$ of the central region of positive vector magnetization correlations, obtained by summing $G_i(\delta \mathbf{\rho})$ over all spin components. The typical domain length $l = \sqrt{4 A/\pi}$ grows continuously from its initial value of about 10 $\mu$m to a final value $l\simeq 40 \, \mu$m over the 3 s equilibration time examined in our measurements (Fig. \ref{fig:domainarea}).  

Phase-ordering kinetics from an initially disordered state have been studied in numerous systems \cite{bray02}. Coarsening dynamics is often found to yield a self-similar domain pattern with a characteristic length scale that evolves temporally as a power law $l(t) \propto t^b$, where the exponent $b$ depends on the order parameter of the system.  Adapting such classical theories to the coarsening of spin textures in a spinor condensate \cite{keekwon2011, josserand1997}, one expects the conserved order parameter to vary effectively from a scalar order parameter (easy axis), for which $b = 1/3$,  to a vector order parameter (easy plane/isotropic), for which $b=1/4$, according to the quadratic Zeeman shift \cite{sieg93,mark95}.  For $q>0$ we expect similar domain coarsening dynamics to the behavior observed for $q=0$, whereas for $q<0$, the rate of domain growth is expected to be slightly faster. 

The observed coarsening dynamics and the ability to tune the order parameter in a spinor condensate should enable a quantitative comparison with theories of phase-ordering kinetics. However, limitations specific to our present experiment prohibit a comparison with theory.  These limitations include the finite size of the system, which when comparable to the domain size may modify the observed coarsening dynamics.

In conclusion, we observe four temporal trends:  a redistribution of Zeeman populations, the development of magnetization from the initially unmagnetized gas, the development of a spin-space anisotropy, and the coarsening of spin domains.  For small $|q|$, all of these trends occur on a similar 1-2 s timescale, allowing the system to reach a common final state independent of the initial spin composition.  For larger $|q|$, equilibrium is not reached within the accessible experimental timescales. The timescale for spin mixing dynamics is set by the spin-dependent interaction energy, $2 |c_2| n \approx h \times 8$ Hz, resulting in spin mixing timescales of  $\tau_{sm}=h/(2 |c_2|  n)$, far shorter than the observed seconds-long equilibration times.  It remains unclear what determines the dynamical timescales for each of the observed processes. 

We thank M. Vengalattore, F. Serwane, and S. R. Leslie  for assistance in the initial design and construction of the experimental apparatus. C. K. Thomas acknowledges support by the Department of Energy Office of Science Graduate Fellowship Program (DOE SCGF), made possible in part by  the American Recovery and Reinvestment Act of 2009, administered by ORISE-ORAU under contract no. DE-AC05-06OR23100. This work was supported by the NSF and by a grant from the Army Research Office with funding from the Defense Advanced Research Projects Agency Optical Lattice Emulator program.\\



\begin{bibliography}{allrefs_x2,equilnotes}

\end{bibliography}


\begin{thebibliography}{34}
\bibitem[{not({\natexlab{a}})}]{note:zeta1}
\bibinfo{note}{From a mean-field treatment of a homogeneous ferromagnetic
  spinor condensate with uniform density $n_0$ we obtain $\zeta_{\pm 1} =
  \left(2 - \epsilon \right)/8$ with $\epsilon = - q / c_n n_0$ in the range $0
  < \epsilon \leq 2$.}
\bibitem[{not({\natexlab{b}})}]{note:variance}
\bibinfo{note}{In practice, to lower the noise in variance measurements, we set
  $G_{L,T}(0)$ equal to the averages of the respective magnetization
  correlation functions over a $3.5 \times 3.5 \, \mu\mbox{m}^2$ region
  centered at $\delta \mathbf{\rho}= 0$.}
\end{thebibliography}
\end{document}